\documentclass[]{svjour3}

\usepackage{graphicx}
\usepackage[round]{natbib}

\journalname{Journal of Molecular Evolution}

\begin{document}

\title{Autocatalytic Sets and RNA Secondary Structure}
\titlerunning{Autocatalytic Sets and RNA Secondary Structure}
\author{Wim Hordijk}
\institute{Wim Hordijk \at
           Konrad Lorenz Institute for Evolution and Cognition Research \\
           Klosterneuburg, Austria \\
           \email{wim@WorldWideWanderings.net}}
\date{Received: date / Accepted: date}
\maketitle

\begin{abstract}
The dominant paradigm in origin of life research is that of an RNA world. However, despite experimental progress towards the spontaneous formation of RNA, the RNA world hypothesis still has its problems. Here, we introduce a novel computational model of chemical reaction networks based on RNA secondary structure and analyze the existence of autocatalytic sub-networks in random instances of this model, by combining two well-established computational tools. Our main results are that (i) autocatalytic sets are highly likely to exist, even for very small reaction networks and short RNA sequences, and (ii) sequence diversity seems to be a more important factor in the formation of autocatalytic sets than sequence length. These findings could shed new light on the probability of the spontaneous emergence of an RNA world as a network of mutually collaborative ribozymes.

\end{abstract}

\section{Introduction}

The dominant paradigm in origin of life research is that of an RNA world \citep{Gilbert:86,Joyce:02}. However, despite experimental progress towards the spontaneous formation of RNA \citep{Powner:09}, the RNA world hypothesis still has its problems \citep{Benner:12,Szostak:12}, and so far no one has been able to show that RNA can catalyze its own template-directed replication.

What has been shown, though, is that some RNA molecules can catalyze the formation of {\it other} RNA molecules from shorter RNA fragments \citep{Horning:16}. Moreover, there are experimentally constructed sets of RNA molecules that {\it mutually} catalyze each other's formation \citep{Sievers:94,Kim:04,Lincoln:09,Vaidya:12}. Rather than each RNA molecule replicating itself, they mutually help each other in being formed from their basic building blocks, in a network of molecular collaboration \citep{Higgs:15,Nghe:15}.

Such a collaborative RNA network is a realization of an {\it autocatalytic set}, a concept that was originally introduced by \citet{Kauffman:71,Kauffman:86,Kauffman:93}. Informally, an autocatalytic set (or RAF set, for {\it R}eflexively {\it A}utocatalytic and {\it F}ood-generated) is a chemical reaction network in which (i) each reaction is catalyzed by at least one molecule from the set itself, and (ii) all molecules can be built up from an appropriate food source through a series of reactions from the set itself. This concept was made mathematically more rigorous and studied in detail, both theoretically and computationally, as RAF theory \citep{Steel:00,Hordijk:04,Mossel:05,Hordijk:17}. This theory has been applied to analyze computational models of chemical reaction networks \citep{Hordijk:13c}, as well as real chemical and biological networks \citep{Hordijk:13a,Sousa:15}.

The computational models that RAF theory has been applied to are mostly variants of a simple polymer model, where molecules are represented by binary strings. In these models, only the primary sequence is taken into account when determining which molecules can catalyze which reactions. However, in real chemistry it is often the secondary (or even tertiary) structure that determines a molecule's catalytic capability. Other (related) computational studies on the emergence and evolution of autocatalytic sets have so far also ignored actual molecular structure \citep{Farmer:86,Bagley:91a,Bagley:91b,Wills:00,Jain:01,Jain:02,Filisetti:11,Vasas:12,Tanaka:14}.

Here, we introduce and analyze a novel model by combining two well-established computational methods: one for predicting RNA secondary structure \citep{Lorenz:11} and one for detecting and analyzing RAF sets \citep{Hordijk:15}. We then study the existence of autocatalytic sets in random instances of this model, where the catalysis assignments are based on RNA secondary structure. Our main result is that autocatalytic sets are highly likely to exist in such systems, even for very small reaction networks and short RNA sequences. Furthermore, this probability increases rapidly with increasing system (network) size and increasing RNA sequence length, but seems to be mostly driven by sequence diversity rather than sequence length. These findings could shed new light on the probability of the spontaneous emergence of an RNA world as a network of collaborative ribozymes.

\section{Methods}

\begin{figure}[htb]
\centering
\includegraphics[scale=0.4]{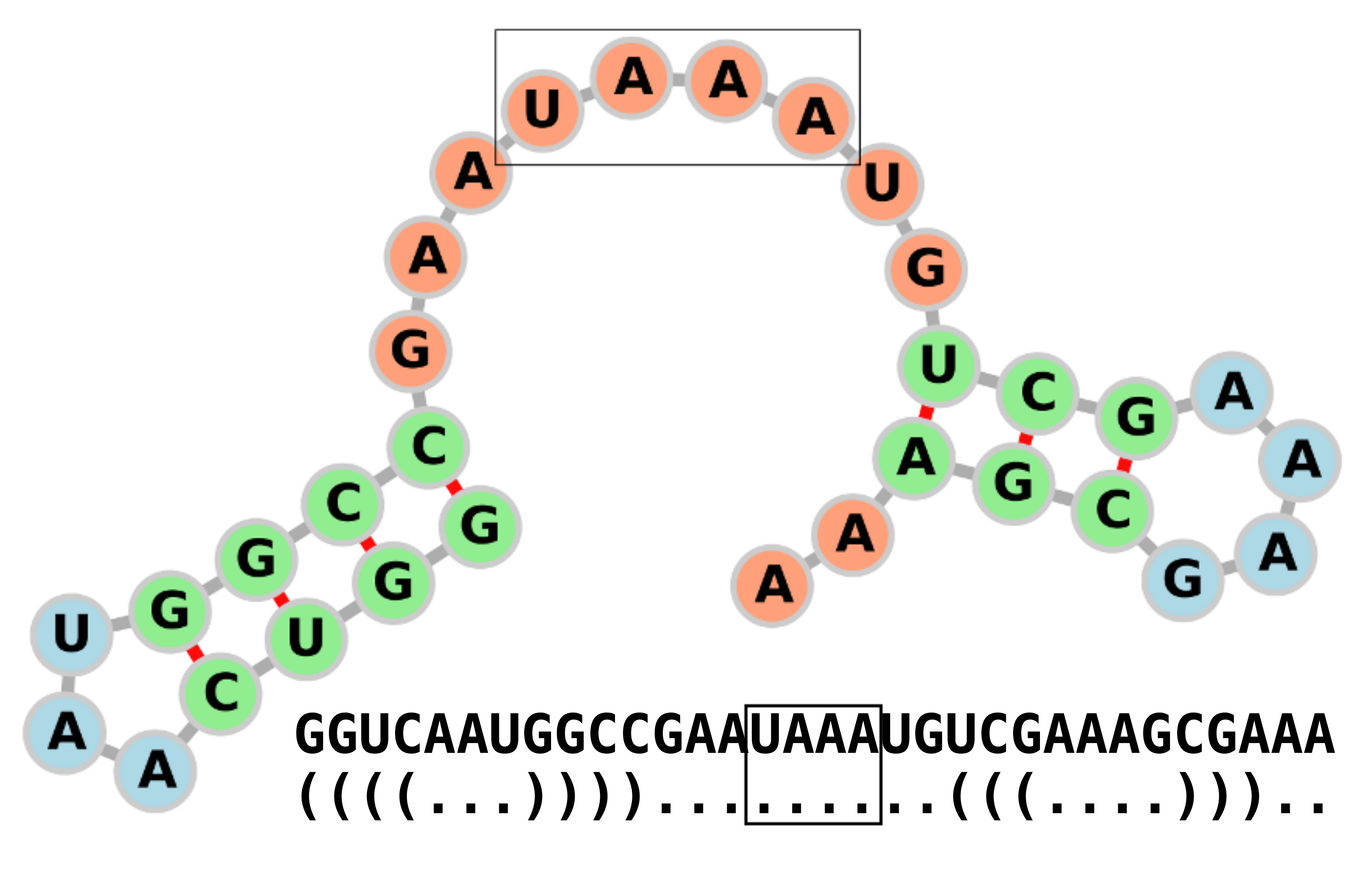}
\caption{An example of a random RNA sequence of length $L=32$, its predicted minimum free energy folded structure (in dot-parentheses notation), and a graphical representation of its secondary structure. Green nucleotides are in stacks (paired), blue ones in hairpin loops (unpaired), and orange ones (unpaired) in neither. The black rectangles indicate the ligation template.}
\label{fig:rna_example}
\end{figure}

We combine two established computational tools to construct and analyze model instances of RNA reaction networks where catalysis is determined by an RNA's secondary structure. First, we generate $N$ random RNA sequences of a given length $L$, where at each sequence position there is an independent and uniform probability of having any of the four nucleotides {\bf A}, {\bf C}, {\bf G}, or {\bf U}. We then use the {\tt ViennaRNA 2.0} package \citep{Lorenz:11} to fold these RNA sequences into their minimum free energy (MFE) secondary structure. This structure is represented using the common dot-parentheses notation. An example is provided in Figure \ref{fig:rna_example} for $L=32$.

Next, we assume that each RNA sequence is broken into two smaller fragments, which can be combined together again through a chemical reaction to (re)form the full sequence. There are at least two such reactions for which empirical support exists: (i) ligation, i.e., a reaction at a phosphoanhydride bond \citep{Bartel:93}, and (ii) recombination, i.e., a reaction at a phosphodiester bond \citep{Hayden:06}. However, in the model an RNA sequence can only be broken at a place along the sequence where there is a consecutive subsequence of at least four unpaired nucleotides (corresponding to at least four consecutive dots in the secondary structure). In particular, we choose that subsequence of four unpaired nucleotides that is closest to the center (mid-point) of the full RNA sequence. The black rectangle in Figure \ref{fig:rna_example} shows this subsequence for the given example, which in this case happens to be exactly at the mid-point of the full sequence. This, then, gives rise to a ``ligation template'' of four nucleotides (two on each side of the ligation site). In the example in Figure \ref{fig:rna_example}, this ligation template is {\bf UAAA}.

Furthermore, for each (full) RNA sequence, we extract the subsequences of all its hairpin loops. In the example in Figure \ref{fig:rna_example} there are two loops (the blue nucleotides), with subsequences {\bf AAU} and {\bf AAAG}.

We now construct an RNA reaction network as follows. The molecule set consists of the $N$ random RNA sequences and their respective fragments (determined by the ligation sites as described above). The reaction set consists of the $N$ ligation reactions that form the full RNA sequences from their respective fragments. Finally, such a ligation reaction can be catalyzed by a full RNA sequence if one of its loops contains a subsequence that is the complementary base-pair match of the ligation template of the first (to be ligated) sequence. For example, the ligation reaction for the RNA sequence shown in Figure \ref{fig:rna_example} (with ligation template {\bf UAAA}) can be catalyzed by another (full) RNA sequence that has a loop containing the subsequence {\bf AUUU}. On the other hand, the example RNA sequence itself (with a loop subsequence {\bf AAAG}) can catalyze any ligation reaction of an RNA sequence with a ligation template {\bf UUUC}. Note that the other loop (of length three) in the example is too short to match a ligation template, and can thus not be used in catalysis.

This novel model is inspired by actual experimental systems that consist of catalytic RNA molecules (ribozymes) which are broken into smaller fragments that can then be joined back together again through a chemical reaction, catalyzed by other ribozymes \citep{Kim:04,Lincoln:09,Vaidya:12}. Moreover, \citet{Lincoln:09} allowed a 4-nt subsequence (at either end of the molecule) to vary, and \citet{Vaidya:12} varied a 3-nt subsequence that acts as the recognition region for catalysis. In our model, we allow for fully random sequences, but use a 4-nt ligation (or ``recognition'') template, in accordance with these experimental systems.

Finally, given a random instance of the RNA reaction network model, we apply the RAF algorithm \citep{Hordijk:15} to detect and analyze the existence of autocatalytic sets, where the RNA fragments are considered to constitute the food source. This process is then repeated 1000 times (for a given set of parameter values $N$ and $L$) to collect statistics on the probability and sizes of autocatalytic sets existing in random instances of the model.

\section{Results}

\begin{figure}[htb]
\centering
\includegraphics[scale=0.6]{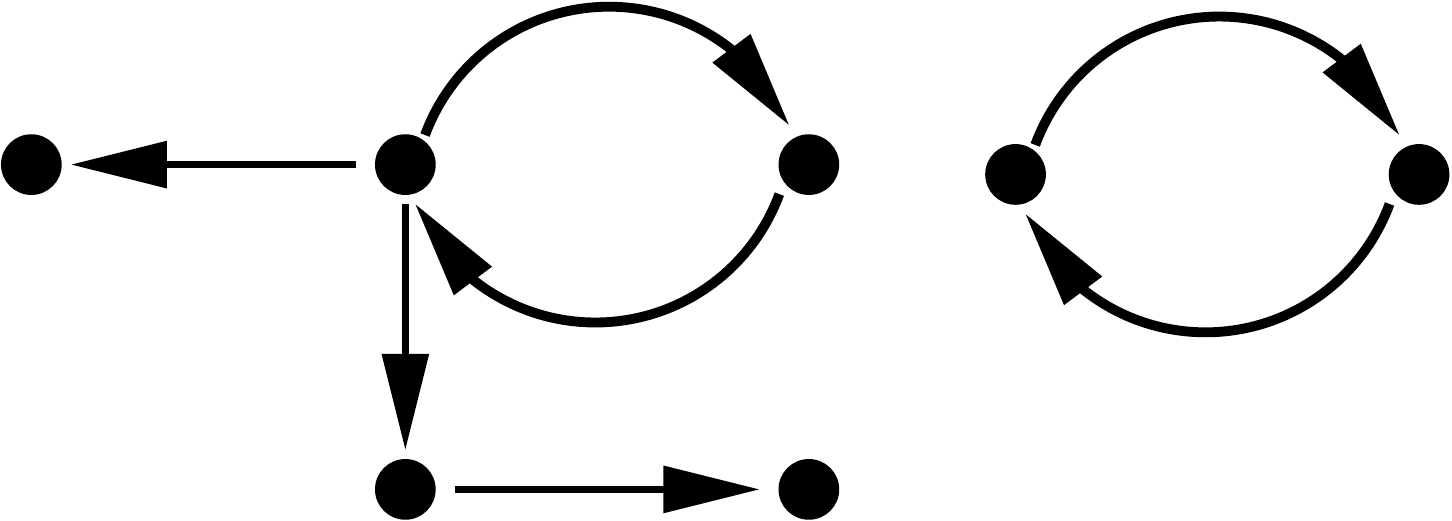}
\caption{An example of an autocatalytic set as found by the RAF algorithm in an instance of the RNA reaction network model with $N=20$ and $L=32$. Dots represent RNA sequences (and their corresponding ligation reaction), and arrows show which sequences catalyze the ligation of which others (as determined by their secondary structure).}
\label{fig:RAF1}
\end{figure}

Taking $N=20$ and $L=32$ as default parameter values, the probability of autocatalytic sets existing in random instances of the RNA reaction network model is about Pr[RAF]=0.5. In other words, even for very small system sizes ($N=20$) and short RNA sequences ($L=32$), about half of the 1000 model instances considered contain a RAF set. Figure \ref{fig:RAF1} shows an example of one such set found by the RAF algorithm. Note that it contains two independent loops (two molecules mutually catalyzing each other's ligation), one of which also catalyzes further members of the set. So, each ligation reaction in this set is catalyzed by one of the molecules from the set, and each molecule in the set is produced from the food source (RNA fragments) through a ligation reaction from the set, thus forming a proper autocatalytic set.

The example shown in Figure \ref{fig:RAF1} contains seven members, i.e., seven of the $N=20$ random RNA sequences catalyze each other's ligation from their respective fragments, in a self-sustaining way. On average the RAF set size is about three (measured over those roughly 500 instances that actually contained a RAF set), with a maximum observed RAF set size of 11. Of course there are still other catalysis events in the RNA network as a whole, i.e., among the RNA molecules that are not included in the RAF set, but they do not contribute to the self-sustaining and catalytically closed autocatalytic set.

\begin{figure}[htb]
\centering
\includegraphics[scale=0.5]{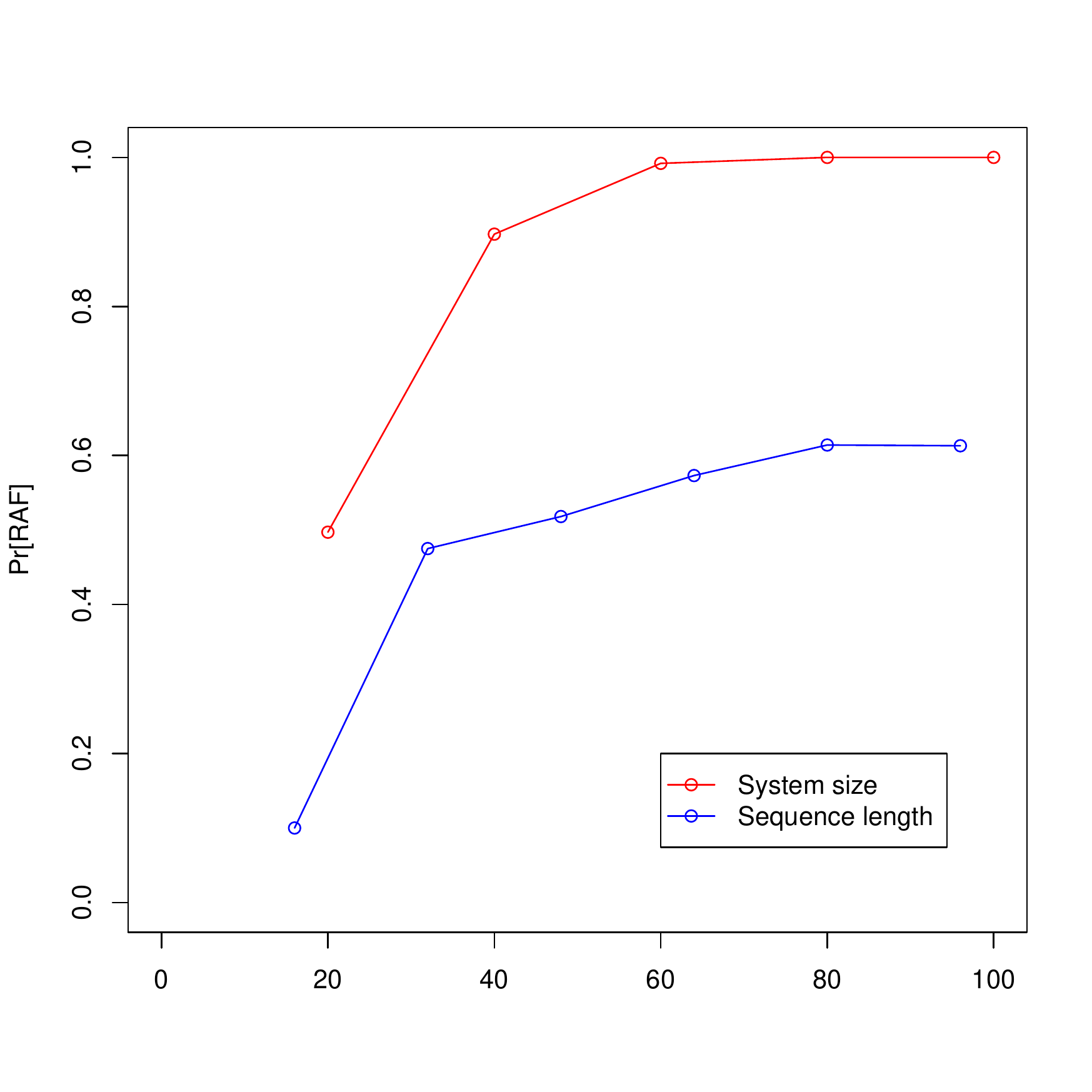}
\caption{The probability of autocatalytic sets (Pr[RAF]) existing in random instances of the RNA reaction network model for different system sizes $N$ (in red, using $L=32$) and sequence lengths $L$ (in blue, using $N=20$), measured over 1000 instances for each parameter value.}
\label{fig:prob}
\end{figure}

The actual probability of a RAF set existing in a random model instance obviously depends on the two parameters $N$ and $L$. Figure \ref{fig:prob} shows these probabilities Pr[RAF] for different values of the system size ($N$, in red, using $L=32$) and sequence length ($L$, in blue, using $N=20$), again measured over 1000 model instances for each parameter value. These probabilities increase rapidly with increasing system size or sequence length. Notably, though, for the short sequence length of $L=16$ and the default system size $N=20$, there is still a 10\% probability that a RAF set exists. In other words, even for the smallest systems and sequences, autocatalytic sets have a non-zero probability of existing, and it does not require a very large increase in either of these parameter values to get autocatalytic sets with a very high likelihood, even among completely random RNA sequences.

\begin{figure}[htb]
\centering
\includegraphics[scale=0.5]{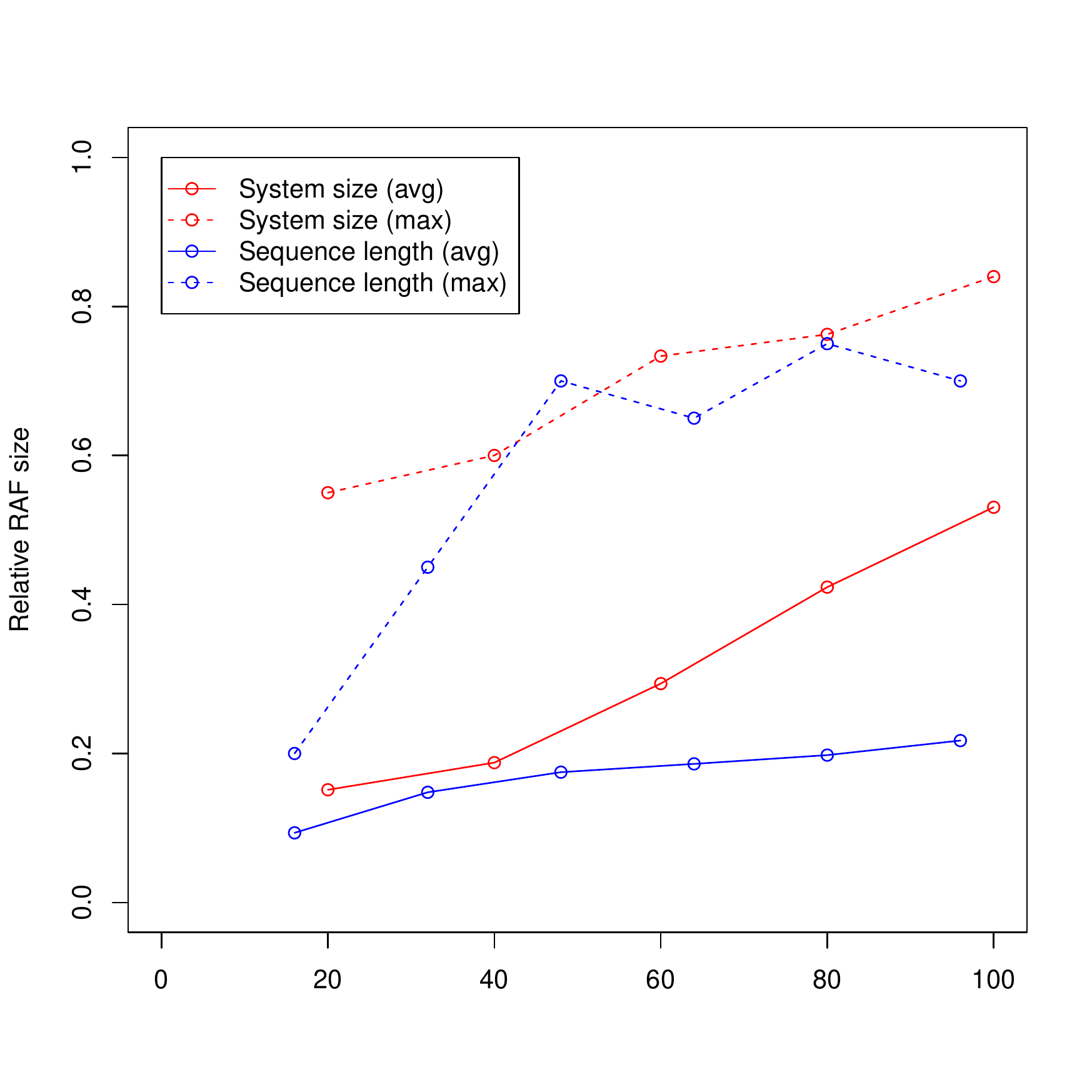}
\caption{The relative sizes of autocatalytic sets (``Relative RAF size'') in random instances of the RNA reaction network model for different system sizes $N$ (in red, using $L=32$) and sequence lengths $L$ (in blue, using $N=20$). Solid lines indicate the average size (relative to the network size) and dashed lines the maximum observed (relative) size, measured over 1000 instances for each parameter value.}
\label{fig:RAFsize}
\end{figure}

Looking at the sizes of the RAF sets, however, there is a difference between the two parameters. Figure \ref{fig:RAFsize} shows the average and maximum relative RAF sizes for different values of the system size ($N$, in red, using $L=32$) and sequence length ($L$, in blue, using $N=20$). These RAF sizes are shown relative to the total number of reactions in the network (i.e., the system size $N$), for a fair comparison between the two parameters (given that the system size $N$ is kept fixed when the sequence length $L$ is varied). Solid lines show the average, and dashed lines the maximum observed.

As the figure shows, the average relative RAF size grows significantly faster with increasing system size $N$ (solid red line) than with increasing sequence length $L$ (solid blue line). Furthermore, the maximum observed relative RAF size continues to increase with increasing system size $N$ (dashed red line), while it appears to level off for increasing sequence length $L$ (dashed blue line). This seems to imply that sequence {\it diversity} (i.e., system size $N$) is a more crucial factor for the existence of autocatalytic sets than sequence {\it length} ($L$).

\begin{figure}[htb]
\centering
\includegraphics[scale=0.5]{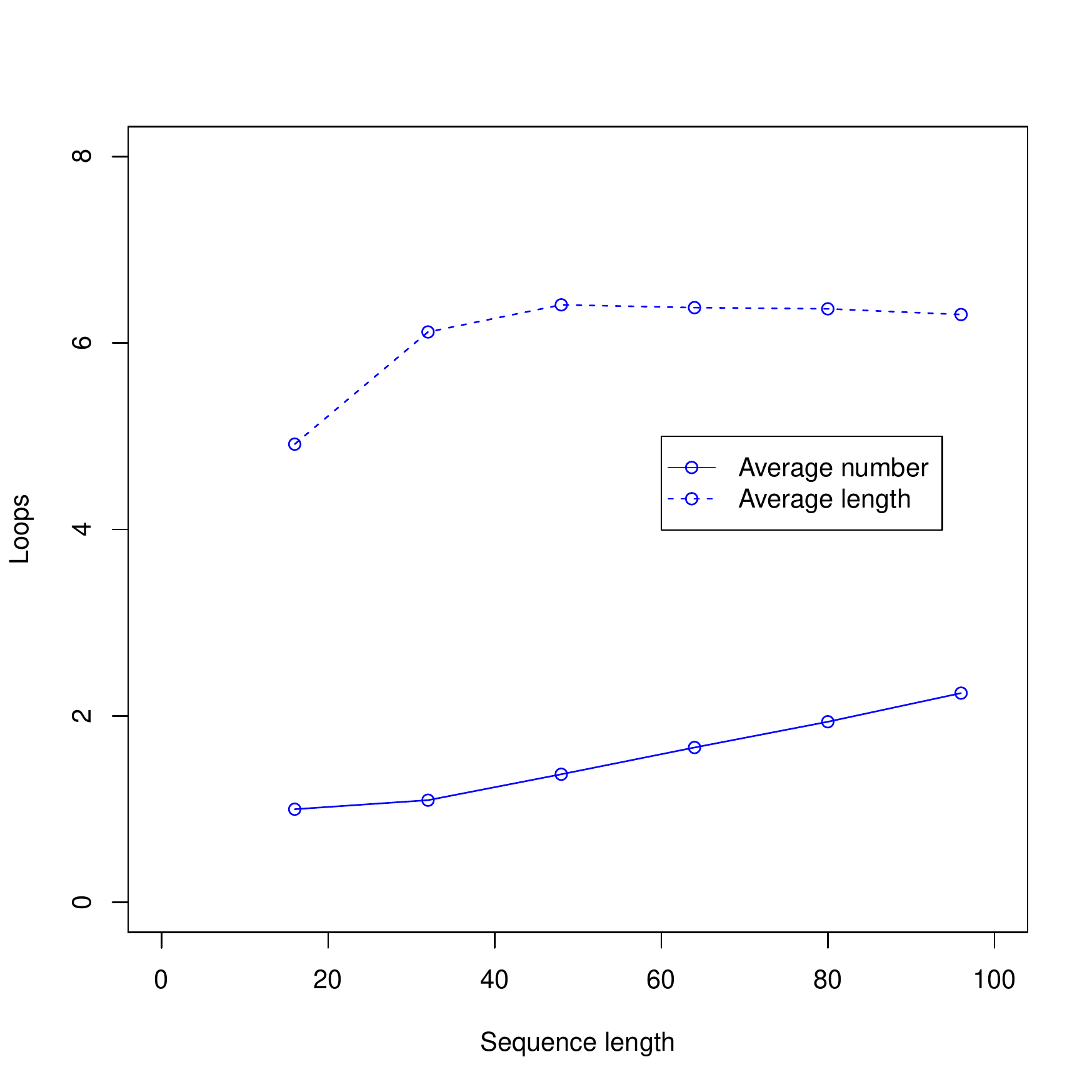}
\caption{The average number of loops (solid line) and their average length in number of nucleotides (dashed line) for random RNA sequences of different sequence lengths $L$, measured over 1000 model instances for each parameter value.}
\label{fig:loops}
\end{figure}

One possible explanation for this can be found in the number and average length of loops in the RNA secondary structures for different sequence lengths. Figure \ref{fig:loops} shows these numbers, with the solid line indicating the average number of loops and the dashed line the average loop length (in number of nucleotides). The average number of loops increases only very slowly with increasing sequence length, and the average loop length plateaus quickly at around six nucleotides. In other words, increasing the sequence length does not necessarily increase a molecule's ability to catalyze more ligation reactions.

\section{Discussion}

The model introduced here is still a simplification of real chemistry, but it is directly inspired by actual experimental RNA systems, and uses an established and reliable method for predicting RNA secondary structure. Moreover, it represents a significant step forward in computational models for studying the existence of autocatalytic sets, by using actual molecular structure to determine catalytic capability, something that is missing in currently existing models.

The main result of this new model is that, even when catalysis is restricted to loops that contain the base-pair complement of a ligation template, determined by an RNA molecule's folded structure, autocatalytic sets are still likely to exist, even in the extreme case of very small networks of short random sequences. Moreover, this probability increases rapidly with increasing system (network) size and RNA sequence length, but seems to be mostly driven by sequence diversity ($N$) rather than sequence length ($L$). These striking results could shed new light on the probability and mechanisms of an RNA world emerging from collections of random RNA sequences, not as individual self-replicators, but as a network of collaborating ribozymes.

Of course many additional features can be added to the basic model. For example, only ligation reactions between the two fragments of a single fully formed RNA molecule have been considered so far. But if these smaller fragments of different RNA sequences are all present in the same solution, they could also (randomly) combine with each other, creating even more possible RNA molecules of length $L$, next to the $N$ ones that were (initially) chosen. However, this would simply create a larger reaction network with more reactions that can potentially be catalyzed by the same molecules. If the more restricted reaction networks generated by the current model already have a high probability of containing a RAF set, these extended networks (of which the more restricted ones are a subset) would have an even higher chance of containing RAF subsets, and possibly even larger ones.

Furthermore, catalysis is currently restricted to hairpin loops in the folded RNA molecules, but could in principle occur in any subsequence that contains at least four unpaired nucleotides. However, this would simply increase the probability that a fully formed RNA molecule catalyzes an arbitrary ligation reaction, thus also resulting in an even higher probability that a model instance contains a RAF set, or possibly an even larger one. What is surprising and encouraging here, is that already in the basic (restricted) model there is such a high probability of observing autocatalytic sets.

As was noted above, the example RAF set in Figure \ref{fig:RAF1} contains two independent autocatalytic sub-networks. For the standard binary polymer model (not taking molecular structure into account), we had already shown that autocatalytic sets usually have a hierarchical structure of smaller and smaller autocatalytic subsets \citep{Hordijk:12}. The current example shows that this is also likely to be the case for the model introduced here, taking RNA secondary structure into account. It has been argued elsewhere that the existence of multiple (independent) autocatalytic subsets within a chemical reaction network is one of the requirements for them to be evolvable \citep{Vasas:12,Hordijk:14}. This requirement thus seems to be fulfilled in our novel model as well.

It is important to note, though, that the current model is not quite representative of a pre-biotic scenario yet, as it assumes the presence of the shorter RNA fragments. However, as mentioned earlier, this novel model is inspired by actual experimental autocatalytic sets that consist of catalytic RNA molecules or peptides which are broken into smaller fragments that can then be joined back together again \citep{Kim:04,Ashkenasy:04,Lincoln:09,Vaidya:12}. Moreover, as the very first experimental evidence for autocatalytic sets has shown \citep{Sievers:94}, this process can even start with simple trimers forming hexamers through mutually catalyzed reactions.

The analysis presented here only focuses on network topology, and does not (yet) take actual dynamics into account. However, in previous simulation studies of experimental RNA autocatalytic networks such dynamical studies were actually performed \citep{Hordijk:13a}, also using experimentally measured reaction rates \citep{Hordijk:14b}. These simulation studies provided more insight into how different RNA autocatalytic sets can come into existence over time, and how environmental influences affect this process. We hope to perform similar dynamical analyses on the model networks introduced here.

Finally, the experimentally verified existence of autocatalytic sets consisting of peptides rather than RNA \citep{Ashkenasy:04}, together with plausible evidence that RNA and peptides interacted and co-evolved very early on in the origin of life \citep{Li:13,Polyanski:13}, would make the formation of one or more autocatalytic sets even more likely, as they increase sequence diversity. In fact, RAF theory is not restricted to RNA molecules alone (or any single type of molecule), and has already been applied to models of ``partitioned'' chemical reaction networks, as with RNA and peptides \citep{Smith:14}. Including even more chemical realism in our novel computational model using actual molecular structure (and the catalytic capabilities determined by it), and also combining different types of molecules, seems a promising direction for learning more about possible routes to the origin of life.

\begin{acknowledgements}
The author thanks the KLI Klosterneuburg for financial support in the form of a fellowship, and two anonymous reviewers for helpful suggestions to improve the original manuscript.
\end{acknowledgements}

\bibliographystyle{spbasic}
\bibliography{RNA_RAF}

\end{document}